\documentclass{article}
\pdfoutput=1
\usepackage{amsmath, amssymb, amsthm}
\usepackage{hyperref}
\usepackage[utf8]{inputenc}
\usepackage[numbers]{natbib}

\title{Intrinsic Regularization via Curved Momentum Space: A Geometric Solution to Divergences in Quantum Field Theory}
\author{Daniel Ketels}
\date{February 20, 2025}

\begin{document}
\maketitle
\begin{abstract}
The problem of ultraviolet (UV) divergences in quantum field theory (QFT) has long been a fundamental challenge. Standard regularization techniques—such as momentum cutoffs, dimensional regularization, and renormalization—modify high-energy behavior to ensure well-defined integrals. However, these approaches often introduce unphysical parameters, rely on arbitrary prescriptions, or break fundamental symmetries, making them mathematically effective but conceptually unsatisfactory.  

In this work, we propose a novel and self-consistent approach in which UV regularization emerges naturally from the geometric structure of momentum space. By introducing a dynamically curved momentum-space metric, we construct an intrinsic measure that automatically suppresses high-energy divergences while preserving fundamental symmetries, including full Lorentz invariance. Unlike traditional methods, this framework requires no explicit cutoffs, does not alter equations of motion, and retains full compatibility with standard field-theoretic formulations.  

This approach ensures the weakest possible suppression necessary for convergence, avoiding excessive modifications to quantum behavior while still achieving finite results. The framework extends seamlessly from a Riemannian formulation to Minkowski space, maintaining its regularization properties in relativistic QFT. Furthermore, it offers a natural alternative to ad hoc renormalization techniques by providing an intrinsic, mathematically well-motivated suppression mechanism rooted in the geometry of momentum space.  

We rigorously construct the measure-theoretic foundations of this framework and demonstrate its effectiveness by proving the finiteness of key QFT integrals. Beyond resolving divergences, this work suggests broader applications in spectral geometry, effective field theory, and potential extensions to quantum gravity, where momentum-space modifications play a fundamental role.
  
\end{abstract}


\section{Introduction}

The problem of ultraviolet (UV) divergences in quantum field theory (QFT) has been a fundamental challenge for decades. Standard regularization methods—such as momentum cutoffs, dimensional regularization, and renormalization techniques—modify the behavior of high-energy modes to make integrals well-defined. However, these methods often involve ad hoc assumptions or artificial modifications that lack a clear theoretical justification from within the structure of the theory itself. For example, lattice discretization explicitly breaks the continuum nature of space-time~\cite{wilson1974}, while dimensional regularization introduces a formally consistent but physically unintuitive analytic continuation of space-time dimensions~\cite{hooft1972}. Similarly, Pauli-Villars regularization modifies propagators by introducing \textit{unphysical ghost fields}~\cite{pauli1949}, which may not emerge naturally from fundamental principles.  

Increasing evidence suggests that momentum space itself may be curved, naturally modifying integration measures at high energies. In particular, studies in deformed special relativity (DSR) have explored curved momentum space geometries as a consequence of quantum gravity effects~\cite{amelino2013, magueijo2002}. Some approaches propose a non-trivial momentum-space metric could arise from the generalized uncertainty principle (GUP)~\cite{maggiore1993} or from relative locality models, where space-time non-commutativity leads to momentum-space curvature~\cite{freidel2011}. These models fundamentally modify the dispersion relation, but they are typically tailored for Planck-scale physics and may not directly apply to conventional QFT.  

This work presents a fundamentally different approach: instead of introducing deformations tied to Planck-scale effects, we propose a geometrically induced measure space in momentum space that emerges from a dynamically curved metric. Our formulation is inspired by the idea that UV divergences should be addressed intrinsically within the mathematical structure of QFT, rather than being imposed externally through additional prescriptions. By modifying the metric structure in momentum space, we construct a natural integration measure that intrinsically suppresses high-energy divergences. Unlike prior approaches that impose momentum cutoffs or modify dispersion relations, our method achieves regularization without altering the fundamental equations of motion or field dynamics. 

Furthermore, while our initial formulation is set in a Riemannian framework, we extend our analysis to Lorentzian space-time, ensuring compatibility with relativistic quantum field theory. This extension addresses challenges related to the negative signature components in the metric and explores the implications for Lorentz invariance and CPT symmetry. Consequently, our method provides a consistent approach to regularization in both Euclidean and Minkowski settings, potentially offering insights for applications in high-energy physics and quantum gravity.

\section{Why Dynamically Curved Momentum Space?}\label{p2}

In traditional quantum field theory, momentum space is assumed to be flat, defining distances and volumes using the standard Euclidean or Minkowski metric. However, this assumption overlooks the possibility that the geometry of momentum space itself could influence field-theoretic behavior at different energy scales. Previous studies, such as deformed special relativity (DSR) \cite{amelino2011}, spectral geometry \cite{connes1994}, and curved-momentum-space approaches in quantum gravity \cite{kowalski2002}, have explored how quantum effects might lead to non-trivial momentum space structures. These works provide the background for our approach, where we investigate a novel application of such geometric modifications in the context of UV divergence suppression.

Inspired by these insights, we propose a framework in which a momentum-dependent metric dynamically modifies distances based on energy scales. Unlike previous approaches focused on Planck-scale effects \cite{amelino2013, magueijo2002}, our formulation is designed to retain Lorentz invariance while suppressing divergences in a minimal and physically motivated manner. Similar to how spacetime curvature emerges from the Einstein equations in General Relativity, our approach treats momentum-space curvature as an intrinsic feature rather than an external modification. The key ingredient is the function $A(p)$, which smoothly interpolates between a nearly flat metric at low momenta and a deformed geometry at high momenta, effectively contracting distances in the ultraviolet (UV) regime. This construction ensures that geodesics, and consequently the measure of integration, adapt dynamically to energy scales, offering an alternative to prior approaches which rely on explicit cutoffs or modified dispersion relations \cite{freidel2011, maggiore1993}.

As a result, high-momentum regions contribute less to quantum field-theoretic calculations, leading to a form of automatic UV regularization. Rather than relying on external cutoffs or dimensional regularization, this framework suggests that intrinsic momentum-space curvature naturally tames divergences. 

\section{Defining an Appropriate Metric Space on \( \mathbb{R}^4 \) }

We consider four-momentum space \( \mathbb{R}^4 \) equipped with a Riemannian metric of a Euclidean signature to ensure positive definiteness. Modified momentum-space geometry has been explored in quantum gravity \cite{momentumspacegravity}, but to the best of the author's knowledge, never in this way. We define the metric tensor as 
\begin{equation} 
g_{\mu\nu}(p) = A(p) \delta_{\mu\nu}, \quad \text{where} \quad A(p) = \frac{1}{1 + \|p-\bar{p}\|^2 \ell_P^2}. 
\end{equation} 
The choice of the reference point \( \bar{p} \in \mathbb{R}^4 \) is arbitrary but fixed, ensuring a well-defined notion of length. As is common in Euclidean geometry, one may conveniently set \( \bar{p} = 0 \).

Thus, within this framework, momentum space is reinterpreted as a four-dimensional Riemannian manifold. \\

\subsection{The Role of the Fundamental Length Scale \( \ell_P \)}

A fundamental length scale \( \ell_P > 0 \) is necessary to define a finite, metric-dependent measure later.It may seem that \( \ell_P \) is simply a parameter of \( A(p) \), yet without it, inherent suppression UV divergences wouldn't work. We will clarify its role now, both geometric and in a dimensional perspective.

\subsubsection{Geometric (Scaling) Properties of \( A(p) \)}

By the metric tensor definition \( g_{\mu\nu}(p) = A(p) \delta_{\mu\nu} \), distance varies only with \( A(p) \), which depends on momentum. Thus, \( A(p) \) scales the metric. Essentially, \( A(p) \) governs the geometry of momentum space. Hence, choosing \( \ell_p \) defines the amount of \textit{curvature}. Only if \( A(p) \) varies does it introduce a more complex geometry, reflecting \textit{energy-dependent curvature} in momentum space. Thus, if \( \ell_P \) is chosen larger, \( A(p) \) approaches a constant more quickly as \( \| p \| \to \infty \), and vice versa. At large momenta, the function behaves asymptotically as
\begin{equation}
    A(p) \sim \frac{1}{\|p\|^2 \ell_P^2}, \quad \text{for} \quad \|p\| \to \infty.
\end{equation}
This ensures that the integration measure decays polynomially at high energies, leading to inherent UV suppression.

\subsubsection{Impact on the Integration Measure}

We will define the measure $\mu$ (an thus the  measure element $d\mu(p)$) by the volume element, to intuitively get the \textit{natural extension of the Lebesgue integral}.\\
As the volume element scales with \( A(p)^2 \), to provides a smooth and intrinsic UV cutoff, the scalar \( \ell_P >0 \) is a necessary part, ensuring that
\begin{itemize}
    \item At low momenta, \( A(p) \approx 1 \), the measure closely retains the Lebesgue measure.
    \item At high momenta, \( A(p) \) decays fast enough, making integrals converge in the UV.
\end{itemize}
As we can see, \( \ell_P \) is central as we define a well-behaved measure space.

\subsubsection{Dimensional Properties: \( \ell_P \) has Natural Energy Scale}

The length scale \( \ell_P \) carries physical units of inverse energy:
\begin{equation}
    [\ell_P] = E^{-1}.
\end{equation}
This follows directly from the requirement that \( A(p) \) must be dimensionless:

momentum has units of energy (\( [p] = E \)), hence the squared norm \( \|p - \bar{p}\|^2 \) has units of \( E^2 \). Thus, \( \ell_P^2 \) has to cancel this unit and theorefore \( \ell_P^2 \) must have units of \( E^{-2} \), i.e. \( \ell_P \) itself must have units of inverse energy.

%Without introducing a fundamental scale like \( \ell_P \), the denominator of \( A(p) \) would lack a natural suppression mechanism for large momenta. This means that the total volume of momentum space could still diverge, and UV regularization would not be automatically built into the theory.

\subsubsection{Connection to Other Theories}

Similar modifications appear in quantum gravity models such as the Generalized Uncertainty Principle \cite{maggiore1993}, where phase-space modifications introduce a minimal length scale. Additionally, non-commutative geometry models \cite{connes1994} suggest that quantum gravity effects lead to deformations in both space-time and momentum-space geometry. Our approach aligns with these ideas, but remains grounded in the standard field-theoretic framework, making it potentially applicable beyond the Planck-scale regime. \\

\noindent Unlike sharp cutoffs or dimensional regularization, our framework guarantees:
\begin{itemize}
    \item Smooth deformation of momentum-space geometry without introducing unphysical parameters.
    \item Preservation of fundamental symmetries such such as Lorentz invariance.
    \item Provides a finite measure without modifying the fundamental equations of motion.
\end{itemize}

Thus, while it is true that \( \ell_P >0 \) simply scales distance in \( A(p) \), it is a necessary and natural part in any well-defined dynamic metric, if the metric is supposed to induce a measure space with some inherentlt regularization of divergences in integrals . \medskip\\

\noindent\textbf{Remark:} One could suspect that $A(p)$ is an artificial choice as a function. The truth is however that it is uniquely induced by the structure of our metric, i.e. the now defined geodesic distance. \textit{In fact, any definition of geodesic distance leads to a metric tensor and vice versa, assuming mild assumptions that are trivially satisfied here. A general proof is given in the apendix}. \\

As this is an initial exploration of metric geometry and measure theory in a physical context, we naturally focus on the physically relevant case of four dimensions. However, the mathematical framework of Riemannian geometry allows a natural extensions to higher-dimensional, if ever necessary.

%Now we define the geodesic distance (which we indirectly did through the metric tensor). The distance is crucial for ensuring that the measure for integration remains well-defined. This emphazies oncemore, that $A(p)$ is not j\textit{ust another suppression function} but rather a fundamental structural element of our curved momentum space.

\subsection{Riemannian Geodesic Distance}
For any distinct \( p,q \in \mathbb{R}^4 \), let \( \Gamma(p,q) \) be the set of all smooth curves \( \gamma:[0,1] \to \mathbb{R}^4 \) such that \( \gamma(0) = p \) and \( \gamma(1) = q \). The Riemannian geodesic distance between the \( p,q \) in \( \mathbb{R}^4 \) is then defined as
\begin{equation}
    d(p,q) := \inf_{\gamma\in\Gamma(p,q)} \left\{ \int_0^1 \sqrt{g_{\mu\nu}(\gamma(s)) \dot{\gamma}^\mu(s) \dot{\gamma}^\nu(s)}\, ds \right\}.
\end{equation}
Since the metric tensor is diagonal, this simplifies to
\begin{equation}
    d(p,q) = \inf_{\gamma\in\Gamma(p,q)} \left\{ \int_0^1 \sqrt{A(\gamma(s))} \| \dot{\gamma}(s) \| ds \right\}.
\end{equation} 
While spectral geometry and path integral modifications have been explored as potential solutions to UV divergences in QFT \cite{spectralgeometry, pathintegraltext}, our work differs by demonstrating how a dynamically curved momentum space metric can naturally induce an integration measure, providing inherent regularization without modifying fundamental equations of motion.

\subsection{Metric Properties of \( (\mathbb{R}^4, d) \)}
To confirm that \( d \) defines a valid metric, we verify its key properties. Non-negativity follows since \( A(p) > 0 \), ensuring \( d(p,q) \geq 0 \), with equality if and only if \( p = q \), since the integral is non-negative (and identically zero for trivial curves). Symmetry is given as any directional change only changes the sign of \( \dot{\gamma} \), which does not change its norm. The triangle inequality follows as any two curves from \( \gamma\in\Gamma_{p,q},\tau\in\Gamma_{q,r} \), combined to a continous $p,r$ path $\gamma\tau$ (generally not differentiable in $q$), cannot exceed the length of a minimal smooth $p,r$ curve.

\begin{equation}
    d(p,r) \leq d(p,q) + d(q,r).
\end{equation}

Since these conditions hold, \( d \) defines a proper metric on \( \mathbb{R}^4 \), making \( (\mathbb{R}^4, d) \) a well-posed metric space. We refer to this space as \textit{dynamically curved momentum space}. \\

We did omit time in the definition for simplicity. Yet this extension is easy, for instance by using an appropriate functional space or by a slightly different definition of $d$ and $\Gamma$. \\
\qed

\subsection{Completeness}

Now we show completeness, i.e. that every Cauchy sequence in \( (\mathbb{R}^4, d) \) converges with limit point in \((\mathbb{R}^4, d) \). We will proceed in three steps:

\paragraph{(i) Comparison with the Euclidean Metric:}
For all \( p \in \mathbb{R}^4 \), we have
\[
A(p) = \frac{1}{1+\|p-\bar{p}\|^2\,\ell_P^2} \leq 1.
\]
Thus, for any smooth curve \( \gamma \in\Gamma(p,q) \), it follows that we have a bound of
\[
\sqrt{A(\gamma(s))} \leq 1,
\]
and by monotonicity of integration:
\[
\int_0^1 \sqrt{A(\gamma(s))} \| \dot{\gamma}(s) \| ds \leq \int_0^1 \| \dot{\gamma}(s) \| ds.
\]
Hence, after taking the infimum of geodesic length, we get
\[
d(p,q) \leq \| p-q \|,
\]
which implies that any Cauchy sequence in \( d \) is also Cauchy in Euclidean standard norm. Since \( \mathbb{R}^4 \) is complete with respect to the Euclidean norm, any Cauchy sequence \( (a_n) \) in \( \mathbb{R}^4 \) must converge to a unique point \( a \in \mathbb{R}^4 \) with $\lim_{n\to\infty} \| a_n -a \|=0$  (or equivalently $a_n\to a$ as $n\to\infty$).

%$\lim_{n\to\infty \| p_n -p \|$ 

\paragraph{(ii) Convergence in the \( d \)-Metric:} First, let us note that \( A \) is continuous on its entire domain and $A(p)$ asymptotically vanishes as \( \|p\| \to \infty \), while $A(p)>0$ (i.e. $A$ remains strictly positive). Thus, for any $p\in\mathbb{R}^4$, there must exist a sufficiently small $\delta>0$ such that the for the $\delta$-Neighborhood \( B_\delta(p) \), in this context defined by the Euclidean norm, a constant \( C > 0 \) exists with 
\[
A(q) \geq C \quad \text{for all } q \in B_\delta(p).
\]
Since \( C > 0 \),  we can choose \( n_0 \in \mathbb{N} \) such that \( \| p_n - p \| < \sqrt{C} \) for all \( n \geq n_0 \). Using the continuity of the infimum, we obtain the inequality
\[
d(p_n, p) \leq \inf_{\gamma\in\Gamma(p_n,p)}\left\{ \int_{0}^1 \sqrt{A(\gamma(s))} \|\dot{\gamma}(s)\| ds \right\} \leq \sqrt{C} \| p_n - p \|.
\]
Since \( \| p_n - p \| \to 0 \) as \( n \to \infty \), it follows that \( d(p_n, p) \to 0 \), proving that convergence in the standard Euclidean norm implies that a sequence is Cauchy in the \( d \)-metric.

\paragraph{Proof Part (iii).} 
One might understandably assume that the proof is complete at this stage. However, a subtle issue remains. In (ii) we showed: \textit{every Cauchy sequence in the} \( d \)\textit{-distance is Cauchy in Euclidean distance.}. But (i) only showed that its \textit{limit point exists in the topology on $\mathbb{R}^4$ induced by the Euclidean distance.} Generally $d$-distance induces a different topology on \( \mathbb{R}^4 \). We have not yet ruled out the possibility that a sequence, which is Cauchy in the \( d \)-metric, could diverge to infinity in the \( d \)-metric. The integral argument in Step (iii) shows: \textit{any path extending to infinite distance in the \( d \)-metric has infinite length, hence no Cauchy sequence can escape to infinity while maintaining finite \( d \)-distance.} Only then can we conclude that the space is complete, as every Cauchy sequence converges \textit{to a point within the space}. If we omit (iii), we could have incompleteness due to missing limit points at infinity.\\

\noindent For \( \|p - \bar{p}\| \to \infty \), we use use asymptotic behaviour of $A$:
\[
A(p) \sim \frac{1}{\|p - \bar{p}\|^2 \ell_P^2}, \quad \text{ and } \quad \sqrt{A(p)} \sim \frac{1}{\|p - \bar{p}\| \ell_P}.
\]
Let $p$ be a radial curve with $p(r) = \bar{p} + r e$, for any unit vector \( e \). Assume \( \| p(r) - \bar{p} \| = r>0 \). As $\dot{p}(r)=\frac{1}{r}$, the length of $p$ from any \( R>0 \) to $\infty$ is
\[
\int_R^\infty \frac{dr}{r \ell_P} = \ell_P \int_R^\infty \frac{dr}{r} = \infty
\]
The integral diverges, hence any sequence escaping to infinity must have infinite \( d \)-distance and therefore is not Cauchy with respect to $d$. This finishes part (iii). \medskip \\ 

\medskip
\noindent \noindent Combining (i), (ii), and (iii), we conclude that every \( d \)-Cauchy sequence converges in \( (\mathbb{R}^4,d) \). Therefore, \( (\mathbb{R}^4, d) \) is a complete metric space, ensuring the formal correctness of our framework. \\ 
\qed

\section{Constructing A Measure Space}\label{p3}

In conventional approaches, ultraviolet (UV) divergences are typically regulated by employing the Lebesgue measure along with external regularization techniques. Traditional renormalization methods can modify integral structures significantly, particularly at multi-loop order, often requiring careful counterterms to ensure finiteness~\cite{collins1984}.\\

In contrast, we define an intrinsic modification of the integration measure by employing the metric we have just constructed. Hence, we define the volume element on \( \mathbb{R}^4 \) as:
\begin{equation}
  \mu(E) = \int_E \sqrt{|\det g_{\mu\nu}(p)|}\, d^4p.
\end{equation}
Since \( g_{\mu\nu}(p) = A(p)\,\delta_{\mu\nu} \) and \( \det(\delta_{\mu\nu}) = 1 \), we have
\[
\det g_{\mu\nu}(p) = A(p)^4, \quad \sqrt{|\det g_{\mu\nu}(p)|} = A(p)^2.
\]
Therefore, our new measure element emerges as
\begin{equation}
  d\mu(p) = A(p)^2\,d^4p.
\end{equation}
Here, \( A(p)^2 \) acts as a suppression factor for large momenta \( p \), effectively regularizing UV divergences. Apart from this weighting factor, the measure retains the structure of the standard Lebesgue integral, ensuring that familiar computational techniques and theorems remain applicable.

\subsection*{The Measure Space \( (\mathbb{R}^4,\mathcal{B},\mu) \)}

Let \( X = \mathbb{R}^4 \) be the momentum space, \( \mathcal{B} \) the Borel \( \sigma \)-algebra on \( \mathbb{R}^4 \), and \( \mu: \mathcal{B} \to [0,\infty] \) the measure induced by the metric \( d \), given by
\[
\mu(E) = \int_E A(p)^2 d^4p.
\]

\paragraph{Non-negativity and \( \sigma \)-additivity:}
Since \( A(p) > 0 \) for all \( p \in \mathbb{R}^4 \), it follows \( A(p)^2 > 0\) and $A(p)^2$ is a measurable function. Thus, for any set \( E \in \mathcal{B} \),
\begin{equation}
\mu(E) = \int_E A(p)^2 d^4p \geq 0.
\end{equation}
The countable additivity of \( \mu \) follows directly from the definition by using the Lebesgue integral properties, ensuring that for any countable  and pairwise disjoint collection \( \{E_k\} \subset \mathcal{B} \),
\begin{equation}
\mu\left( \bigcup_{k=1}^{\infty} E_k \right) = \sum_{k=1}^{\infty} \mu(E_k).
\end{equation}
Hence, we know that \( \mu \) is a well-defined measure on \( (\mathbb{R}^4, \mathcal{B}) \).

\paragraph{\( \sigma \)-Finiteness of \( \mu \):}
A measure \( \mu \) is \( \sigma \)-finite if there exists a countable collection of measurable sets \( \{E_n\} \subset \mathcal{B} \) satisfying
\begin{equation}
X = \bigcup_{n=1}^{\infty} E_n, \quad \text{with} \quad \mu(E_n) < \infty \quad \forall n \in \mathbb{N}.
\end{equation}
To verify this, consider the sequence of closed balls
\begin{equation}
B_n(0) = \{ p \in \mathbb{R}^4 \mid \|p\| \leq n \}, \quad n \in \mathbb{N}.
\end{equation}
Clearly, \( \mathbb{R}^4 = \bigcup_{n=1}^{\infty} B_n(0) \). Moreover, since \( A(p) \leq 1 \), we obtain an upper bound for \( \mu(B_n(0)) \) using the Lebesgue measure:
\begin{equation}
\mu(B_n(0)) = \int_{B_n(0)} A(p)^2 d^4p \leq \int_{B_n(0)} d^4p = \operatorname{Vol}(B_n(0)) < \infty.
\end{equation}
Hence, we conclude that \( \mu \) is \( \sigma \)-finite, as required.\\
 \qed

\section{Minimal Suppression in the Worst Case}\label{p4}

To demonstrate that the function \( A(p)^2 \) provides one of the weakest possible smooth suppressions of ultraviolet (UV) divergences, we first compare it to common damping functions used in regularization~\cite{peskin1995, hooft1972}. Here, "weak" refers to the minimal order of suppression required to ensure convergence, rather than the effectiveness of the suppression itself.

Several well-known functions satisfy the necessary UV regularization criteria. Among these, Gaussian suppression is frequently employed in momentum-space cutoffs~\cite{cutoffQFT}. However, in contrast to these conventional approaches, \( A(p)^2 \) provides the weakest polynomial suppression that still ensures convergence, making it a minimal smooth regulator in this sense.\\

This stands in contrast to methods such as dimensional regularization~\cite{dimreg1972} and lattice discretization~\cite{latticeQFT}, which impose significantly stronger modifications on integrals. More critically, these approaches fundamentally alter both the topological and geometric properties of the governing space. While they can yield useful results, a fundamental theory of physics should ideally avoid unnecessary transformations. This is particularly relevant for discretization techniques, which inherently contradict the continuous nature of field theories.\\

\noindent A smooth function $f$ serving as a UV regulator should satisfy four key properties:  
(1) \textbf{Low-energy consistency}, ensuring \( f(p) \approx 1 \) for small \( p \), preserving low-energy physics.  
(2) \textbf{High-energy suppression}, requiring \( f(p) \to 0 \) as \( p \to \infty \), guaranteeing UV finiteness.  
(3) \textbf{Smoothness}, i.e. \( f \) is continuously differentiable on the complete domain.  
(4) \textbf{Minimal modification}, imposing only the weakest necessary suppression for convergence.  

Several functions meet these criteria. A common example is the Gaussian: \( f(p) = e^{-p^2 \Lambda^2} \) with exponential suppression, making it highly effective but overly strong for mild modifications. A more controlled alternative is power-law suppression, \( f(p) = (1 + p^2/\Lambda^2)^{-n} \), where \( n \) determines the suppression strength. The proposed function, \( A(p)^2 = (1 + \|p - \bar{p}\|^2 \ell_P^2)^{-2} \), provides polynomial suppression and represents the weakest decay still ensuring UV convergence.\\

Comparing their asymptotic behavior as \( ||p|| \to\infty \), the Gaussian regulator decays exponentially as \( e^{-p^2} \). Power-law functions behave as \( p^{-4} \) for \( n = 2 \) (the lowest feasible $n$) and decay more rapidly for \( n \geq 3 \). The proposed function satisfies \( A(p)^2 \sim p^{-4} \), making it of weakest polynomial suppression Order that guarantees convergence.\\

It is important to clarify that this does not represent the worst possible UV divergence in quantum field theory. In renormalizable QFTs, divergences typically scale as \( p^{-4} \), making \( A(p)^2 \sim p^{-4} \) a natural choice for suppression. However, non-renormalizable theories, such as higher-derivative gravity or exotic gauge interactions, often require stronger suppression. Similarly, multi-loop diagrams can exhibit more severe divergences than simple power counting suggests, necessitating stricter regulators. Thus, while \( A(p)^2 \) provides the minimal suppression required for standard renormalizable QFTs, it may not suffice for extreme UV behaviors in certain quantum theories. \\

The following theorem \textit{does not claim that \( A(p)^2 \) is the optimal suppression function in a general sense}, but rather that it provides the weakest suppression necessary to ensure convergence, making it a minimal but sufficient choice.

\subsection*{Minimal Polynomial Suppression by \( A(p)^2 \) for Convergence}

We analyze the behavior of a typical QFT integral without assuming a fixed \( \alpha \):
\begin{equation}
I = \int_{\mathbb{R}^4} \frac{A(p)^2}{(p^2+m^2)^\alpha} d^4p.
\end{equation}
Since \( A(p)^2 \sim p^{-4} \) at large \( p \), we require \( (p^2 + m^2)^{-\alpha} \sim p^{-2\alpha} \) to decay at least as fast as \( p^{-4} \). This condition holds if and only if
\begin{equation}
2\alpha > 0, \quad \text{or equivalently,} \quad \alpha > 0.
\end{equation}

This result parallels the behavior of the generalized harmonic series, which converges only for exponents \( n^\alpha \) with \( \alpha < -1 \).

Hence, in our integral case, we have shown that for any fixed \( \alpha > 0 \), the function \( A(p)^2 \) provides sufficient suppression to guarantee convergence, while remaining the weakest-order smooth regulator that ensures this. \\
\qed

\subsection{Conclusion of Suppression Comparison}

Among all smooth suppression functions ensuring UV convergence, \( A(p)^2 \) is a possible minimal choice, providing the slowest polynomial decay necessary for regularization. By contrast, Gaussian suppression or higher power-law functions impose significantly stronger constraints, altering the structure of integrals more dramatically. \\

However, as mentioned earlier, \( A(p)^2 \) is not necessarily sufficient for all cases. While it ensures convergence in most standard QFT integrals, higher-order divergences in more exotic theories (e.g., quantum gravity, higher-loop diagrams, or non-renormalizable interactions) may require stronger suppression.  ~\cite{collins1984}. \\

In higher-derivative quantum gravity, where divergences are worse than those found in renormalizable QFTs, standard suppression methods may be insufficient~\cite{stelle1977}. Furthermore, power counting in effective field theory suggests that gravity introduces divergences that require non-trivial suppression~\cite{weinberg1979}. \\

In contrast, \( A(p)^2 \) introduces only the minimal required suppression at large \( p \), preserving as much of the original integral’s contribution to the field theory as possible while still ensuring finiteness at high energies.\\

In the next section, we analyze a related convergence problem, demonstrating how the convergence behaviour of QFT integrals can change when using the measure element $d\mu(p)=A(p)^2d^4p$. 

This further reinforces the claim that while \( A(p)^2 \) is as weak as possible, it remains as strong as necessary, excluding extreme scenarios as discussed. In this demonstration, we will see that the integral still converges  if all free parameters assume worst case values, which would typically lead to divergence.

\section{Convergence of a Momentum--Space Integral}\label{p5}

In standard quantum field theory (QFT), integrals of the form
\begin{equation}
I_{\text{QFT}} = \int_{\mathbb{R}^4} \frac{1}{(p^2+m^2)^\alpha} d^4p
\end{equation}
naturally arise, particularly in loop calculations involving the Feynman propagator. However, it is well known that this integral \textit{diverges} for certain values of \( \alpha \), requiring regularization techniques such as dimensional regularization~\cite{dimreg1972} or momentum cutoffs~\cite{cutoffQFT}.

Instead of imposing an external regularization, we reevaluate the integral within the measure space \( (\mathbb{R}^4, \mathcal{B}, \mu) \) constructed earlier:
\begin{equation}
I = \int_{\mathbb{R}^4} \frac{1}{(p^2+m^2)^\alpha} d\mu(p),
\end{equation}

where the measure element is given by
\begin{equation}
d\mu(p) = A(p)^2 d^4p.
\end{equation}

Here, the suppression function \( A(p)^2 \) is not introduced as an arbitary external regulator. Instead, it \textit{emerges naturally} from the geometry of the measure space. This fundamental difference ensures that high-energy contributions are \textit{intrinsically suppressed} without explicitly modifying the integral’s structure.\\

We will now rigorously prove that this integralc is finite \textit{for all choices of \( \alpha > 0, m>0 \)}, demonstrating how the geometric structure of the measure itself introduces enough suppression to regularize the divergence without imposing an artificial cutoff.

\subsubsection*{Proof Strategy}

To establish convergence, we analyze the integral over two disjoint sets:
\begin{enumerate}
    \item \textbf{Bounded region}: \( B_R(0) = \{ p \in \mathbb{R}^4 \mid d(0,p) \leq R \} \).
    \item \textbf{Unbounded region}: \( E_R := \mathbb{R}^4 \setminus B_R(0) \), where we need to verify that \( A(p)^2 \) decays sufficiently fast.
\end{enumerate}
An argument for the \textit{cutoff method} naturally emerges here. Given that \( B_R(0) \) can be arbitrarily large, we can capture as much of the integral as we deem necessary. However, the crucial counterargument is: \textit{how much of the actual integral is sufficient, and how do we determine that?} 

Rather than debating this, we now prove that the integral is finite over all of \( \mathbb{R}^4 \) in our measure.

\subsubsection*{Convergence in the Compact Region}

Since \( B_R(0) \) is bounded and all boundary points \( p \) with \( d(0,p)=R \) are included, \( B_R(0) \) is a compact set in \( \mathbb{R}^4 \). Furthermore, since \( A(p) \leq 1 \) and
\begin{equation}
f(p) = \frac{1}{(p^2+m^2)^\alpha}
\end{equation}
is uniformly continuous on the compact set \( B_R(0) \), the integral
\begin{equation}
\int_{B_R(0)} \frac{A(p)^2}{(p^2+m^2)^\alpha} d^4p \leq \int_{B_R(0)} \frac{1}{(p^2+m^2)^\alpha} d^4p
\end{equation}
is finite. Thus, the integral converges, as it is finite and continuous a compact domain.

\subsubsection*{Asymptotic Behavior in the Large-Momentum Region}

For the unbounded region, we observe that for sufficiently large radius \( R \) (since \( \bar{p} \) is fixed), there must exist a constant \( c > 0 \) such that
\begin{equation}
\|p-\bar{p}\| \geq c \|p\|.
\end{equation}
This provides the bound
\begin{equation}
A(p) \leq \frac{1}{\|p-\bar{p}\|^2 \ell_P^2} \leq \frac{1}{c^2 \ell_P^2 \|p\|^2}.
\end{equation}
Squaring both sides, we obtain
\begin{equation}
A(p)^2 \leq \frac{1}{c^4 \ell_P^4 \|p\|^4}.
\end{equation}
Since \( p^2 + m^2 \geq m^2 > 0 \), we also have
\begin{equation}
\frac{1}{(p^2+m^2)^\alpha} \leq \frac{1}{\|p\|^{2\alpha}}.
\end{equation}
Thus, for any \( p \in E_R \),
\begin{equation}
\frac{A(p)^2}{(p^2+m^2)^\alpha} \leq \frac{1}{c^4 \ell_P^4 \|p\|^{4+2\alpha}}. \vspace{1ex}
\end{equation} 
Switching to spherical coordinates in \( \mathbb{R}^4 \), using \( d^4p = \omega_3 r^3 dr \), the integral satisfies
\begin{equation}
\int_{E_R} \frac{A(p)^2}{(p^2+m^2)^\alpha} d^4p = \frac{\omega_3}{c^4 \ell_P^4} \int_R^\infty \frac{r^3 dr}{r^{4+2\alpha}}.
\end{equation}
Now, this integral is known to converge if and only if \( 1 + 2\alpha > 1 \), or equivalently,
\begin{equation}
\alpha > 0.
\end{equation}
As we have assumed \( \alpha > 0 \), we conclude that the integral converges to a finite value of:
\begin{equation}
\int_{\mathbb{R}^4} \frac{A(p)^2}{(p^2+m^2)^\alpha} d^4p = \int_{B_R(0)} \frac{A(p)^2}{(p^2+m^2)^\alpha} d^4p  + \int_{E_R} \frac{A(p)^2}{(p^2+m^2)^\alpha} d^4p
\end{equation} \\
\qed

\subsection*{Conclusion}

It is crucial that the integral
\begin{equation}
I = \int_{\mathbb{R}^4} \frac{1}{(p^2+m^2)^\alpha} d\mu(p)
\end{equation}
remains convergent \textit{for all} \( \alpha > 0, m>0 \), in contrasts to the original integral \( I_{\text{QFT}} \), which diverges unless \( \alpha > \frac{3}{2} \). \\

Thus, this demonstrates that the suppression provided by \( A(p)^2 \) is \textit{as weak as possible} yet \textit{as strong as necessary}, ensuring convergence without artificially modifying the integral’s structure. Crucially, this suppression is not an externally imposed regularization but an intrinsic consequence of the modified momentum-space geometry. \\

This result underscores the effectiveness of our approach to UV divergence suppression, as the function \( A(p)^2 \) emerges naturally from a constructed metric space \( (\mathbb{R}^4, d) \), thereby validating its use in QFT regularization

\section{Invariance Properties of the Metric and Measure}\label{p6}

In this section, we examine the invariance properties of the metric and measure under fundamental transformations in momentum space. We determine whether invariance holds, and if not, under what conditions it can be restored.

\subsection{Invariance Under Translations in Momentum Space}
\textit{
The modified metric tensor \( g_{\mu\nu}(p) = A(p) \delta_{\mu\nu} \) and the corresponding measure \( d\mu(p) = A(p)^2 d^4p \) are invariant under translations \( p \mapsto p + c \) (for any constant shift \( c \in \mathbb{R}^4 \)) if and only if the reference point \( \bar{p} \) is also shifted accordingly.
}

\begin{proof}
Consider a translation \( p \mapsto p + c \). The suppression function \( A(p) \) transforms as follows:
\begin{equation}
A(p+c) = \frac{1}{1 + \|(p+c)-\bar{p}\|^2 \ell_P^2}.
\end{equation}
If the reference point \( \bar{p} \) is also shifted by \( c \), i.e., \( \bar{p} \mapsto \bar{p} + c \), then:
\begin{equation}
A(p+c) = \frac{1}{1 + \|(p+c) - (\bar{p} + c)\|^2 \ell_P^2} = \frac{1}{1 + \|p - \bar{p}\|^2 \ell_P^2} = A(p)
\end{equation}
Since \( g_{\mu\nu}(p) \) and \( d\mu(p) \) both depend only on \( A(p) \), it follows that they remain unchanged under translations if and only if \( \bar{p} \) is translated accordingly:
\begin{equation}
d\mu(p+c) = A(p+c)^2 d^4(p+c) = A(p)^2 d^4p = d\mu(p).
\end{equation}
However, if for some reason \( \bar{p} \) remains fixed, then:
\begin{equation}
A(p+c) \neq A(p),
\end{equation}
which explicitly breaks translation invariance. Therefore, the measure remains translation-invariant if and only if \( \bar{p} \) is shifted in the same manner as \( p \), i.e., \( \bar{p} \mapsto \bar{p} + c \) (which should generally be the case for a translation). \\
\end{proof}

\subsection{Rotational Invariance}

\textit{
The metric tensor \( g_{\mu\nu}(p) \) and measure \( d\mu(p) \) are invariant under global rotations in Riemannian space with respect to the orthogonal group \( O(4) \).
}

\begin{proof}
Consider a rotation \( p \mapsto \Lambda p \), where \( \Lambda \in O(4) \). The Riemannian norm is preserved under rotations, so
\begin{equation}
\| \Lambda p - \bar{p} \| = \| p - \Lambda^{-1} \bar{p} \|.
\end{equation}
Thus, the suppression function transforms as:
\begin{equation}
A(\Lambda p) = \frac{1}{1 + \| \Lambda p - \bar{p} \|^2 \ell_P^2} = \frac{1}{1 + \| p - \Lambda^{-1} \bar{p} \|^2 \ell_P^2}.
\end{equation}
If \( \bar{p} \) is rotated accordingly as \( \bar{p} \mapsto \Lambda \bar{p} \), then:
\begin{equation}
A(\Lambda p) = A(p).
\end{equation}
Since the measure depends only on \( A(p)^2 \), which remains invariant, the measure is also preserved under such rotations.\\
\end{proof}

\subsection{Scaling Properties}

\textit{
The metric and measure exhibit a well-defined scaling behavior under \( p \to \lambda p \), where \( \lambda > 0 \).
}
\begin{proof}
Under a scaling transformation,
\begin{equation}
A(\lambda p) = \frac{1}{1 + \| \lambda p - \bar{p} \|^2 \ell_P^2}.
\end{equation}
For general \( \bar{p} \), this transformation modifies the functional form of \( A(p) \) unless \( \bar{p} \) is also scaled as \( \bar{p} \to \lambda \bar{p} \). In that case, we obtain:
\begin{equation}
A(\lambda p) = \frac{1}{1 + \lambda^2 \| p - \bar{p} \|^2 \ell_P^2}.
\end{equation}
The measure transforms as:
\begin{equation}
d\mu(\lambda p) = A(\lambda p)^2 d^4(\lambda p) = \lambda^4 A(\lambda p)^2 d^4p.
\end{equation}
Thus, the measure acquires an overall scaling factor,
\begin{equation}
d\mu(\lambda p) = \lambda^4 d\mu(p).
\end{equation}
Hence, while the functional form of \( A(p) \) is altered unless \( \bar{p} \) scales accordingly, the measure retains a well-defined scaling behavior, matching the scaling properties of the Lebesgue measure in four dimensions. \\

\end{proof}

\subsection{Lorentz Invariance in Minkowski Space}

\textit{ If the metric is formulated in Minkowski space, the suppression function \( A(p) \) remains Lorentz-invariant:
\begin{equation}
A(p) = \frac{1}{1 + (p^\mu - \bar{p}^\mu)(p_\mu - \bar{p}_\mu) \ell_P^2}.
\end{equation}
That is, for any \( p \in \mathbb{R}^4 \) and any Lorentz transformation \( \Lambda^\mu_{\ \nu} \),
\begin{equation}
A(\Lambda p) = A(p)
\end{equation}
}

\begin{proof}
The quantity inside \( A(p) \) is the squared Minkowski distance between \( p^\mu \) and \( \bar{p}^\mu \), given by:
\begin{equation}
S(p, \bar{p}) = \eta_{\mu\nu} (p^\mu - \bar{p}^\mu)(p^\nu - \bar{p}^\nu),
\end{equation}
where the Minkowski metric \( \eta_{\mu\nu} \) is
\begin{equation}
\eta_{\mu\nu} =
\begin{bmatrix}
1 & 0 & 0 & 0 \\
0 & -1 & 0 & 0 \\
0 & 0 & -1 & 0 \\
0 & 0 & 0 & -1
\end{bmatrix}.
\end{equation}
Thus, we can rewrite \( S(p, \bar{p}) \) as:
\begin{equation}
S(p, \bar{p}) = (p^0 - \bar{p}^0)^2 - (\mathbf{p} - \bar{\mathbf{p}})^2.
\end{equation}
Under a Lorentz transformation \( p^\mu \mapsto p'^\mu = \Lambda^\mu_{\ \nu} p^\nu \), we analyze the transformation of \( S(p, \bar{p}) \):
\begin{equation}
S'(p', \bar{p}') = \eta_{\mu\nu} (p'^\mu - \bar{p'}^\mu)(p'^\nu - \bar{p'}^\nu).
\end{equation}
Since Lorentz transformations preserve the Minkowski norm,
\begin{equation}
\eta_{\rho\sigma} \Lambda^\rho_{\ \mu} \Lambda^\sigma_{\ \nu} = \eta_{\mu\nu},
\end{equation}
it follows that:
\begin{equation}
S'(p', \bar{p}') = S(p, \bar{p}).
\end{equation}
Substituting this result into \( A(p) \), we obtain:
\begin{equation}
A(p') = \frac{1}{1 + S'(p', \bar{p}') \ell_P^2} = \frac{1}{1 + S(p, \bar{p}) \ell_P^2} = A(p).
\end{equation}
Thus, \( A(p) \) remains exactly Lorentz-invariant.\\

\noindent Since the measure element is
\begin{equation}
d\mu(p) = A(p)^2 d^4p,
\end{equation}
and we established \( A(p') = A(p) \), the measure transforms as:
\begin{equation}
d\mu(p') = A(p')^2 d^4p' = A(p)^2 \det(\Lambda) d^4p.
\end{equation}
For proper Lorentz transformations, \( \det(\Lambda) = 1 \), ensuring full Lorentz invariance. However, for improper Lorentz transformations (which include parity inversion), \( \det(\Lambda) = -1 \) leads to a sign flip in the volume element. This subtle distinction is important when considering CPT-related extensions.\\
\end{proof}

\section{Full Extension to Minkowski Space}

The results above demonstrate that the suppression function \( A(p) \) remains exactly Lorentz-invariant when formulated in Minkowski space.

This suggests that the curved momentum-space approach introduced here is compatible with relativistic quantum field theory (QFT). \\

While our initial formulation was in Riemannian space, the Minkowski extension is important for applications in QFT, where integrals are typically performed in a space-time with signature \( (+,-,-,-) \). Since ultraviolet divergences in QFT arise in Minkowski-space integrals, extending our measure formulation to this setting is essential for practical applications. \\

A key distinction between Euclidean and Minkowski space is that the volume element of the integration measure must correctly reflect the determinant of the metric. However, since the measure element transforms under Lorentz transformations as: 
\begin{equation} d\mu(p') = A(p')^2 d^4p' = A(p)^2 \det(\Lambda) d^4p, 
\end{equation} 
and we have shown that \( A(p) \) itself remains invariant under Lorentz transformations, the measure retains its fundamental structure.\\ 

Hence, our approach is directly applicable in Minkowski space, ensuring that: \begin{enumerate} \item The same measure structure can be used in conventional QFT calculations. \item The method preserves Lorentz invariance, ensuring compatibility with relativistic theories. \item Unlike hard cutoffs or dimensional regularization, this formulation provides a smooth, intrinsic suppression of UV divergences without modifying fundamental equations of motion. \end{enumerate}

Furthermore, since improper Lorentz transformations introduce a sign change in the volume element, i.e. \( \det(\Lambda) = -1 \), future work could explore potential implications for CPT symmetry and its extensions in modified momentum-space geometries.

\subsection{Implications and Constraints}

The fact that \(A(p) \) remains invariant under Lorentz transformations ensures that our modified integration measure is fully compatible with relativistic quantum field theory. However, the sign change in the volume element under improper Lorentz transformations, such as parity inversion ($P$) or time-reversal ($P$), suggests interesting physical consequences. Since CPT symmetry is a cornerstone of quantum field theory, any modification of the measure under discrete transformations could have implications for fundamental symmetry-breaking scenarios~\cite{greenberg2002}.\\

Additionally, in approaches to quantum gravity where the structure of momentum space is modified~\cite{kowalski2002, amelino2011} may benefit from an explicit measure-based approach to controlling divergences while preserving Lorentz symmetry. \\

Another interesting direction involves higher-derivative quantum gravity models~\cite{stelle1977}, where modifications to the propagator introduce power-law suppressed UV divergences that might be tamed within a momentum-space curvature framework. \\

Future work could explore how our framework behaves in curved spacetime and under general diffeomorphisms, potentially linking it to gravitational effective field theories.

\section{Exemplified Application to Quantum Field Theory}\label{p7}

Consider the Euclidean generating functional for a free scalar field \(\phi\) of mass \(m\). In momentum space, define
\begin{equation}
  Z[J] = \int \mathcal{D}\phi \, \exp\Biggl\{ - \int_{\mathbb{R}^4} d\mu(p) \Bigl[ \frac{1}{2}\phi(p)(p^2+m^2)\phi(-p) - J(p)\phi(-p) \Bigr] \Biggr\}.
\end{equation}
The integration measure $\mathcal{D}\phi$ in this case accounts for the modified momentum-space volume element. Since the action is quadratic, the path integral is Gaussian, following the standard formulation of QFT path integrals~\cite{peskin1995, pathintegraltext}.
\begin{equation}
  Z[J] = Z[0]\, \exp\left\{ \frac{1}{2} \int_{\mathbb{R}^4} d\mu(p) \, \frac{J(p)J(-p)}{p^2+m^2} \right\}.
\end{equation}
Taking functional derivatives with respect to $J$ gives the two-point function (propagator), which follows the conventional Feynman propagator approach~\cite{peskin1995}.
\begin{equation}
\langle \phi(p)\phi(-p)\rangle = \frac{1}{p^2+m^2}
\end{equation}

While this result retains the usual form, it is important to note that the modified integration measure \( d\mu(p) \) affects loop integrals by suppressing high-momentum contributions, thereby influencing renormalization. 
Consequently, while tree-level computations remain identical to those in standard QFT, loop corrections - such as self-energy corrections and vacuum polarization - experience significant modification due to the built-in UV suppression in the measure. This affects the renormalization procedure, potentially reducing or even eliminating the need for counterterms in certain cases, as the measure intrinsically regularizes divergent integrals~\cite{hooft1972, polchinski1984}.\\

For reference, the modified measure element is explicitly given by:
\begin{equation}
d\mu(p) = A(p)^2 \frac{d^4p}{(2\pi)^4},
\end{equation}
where \( A(p)^2 \) decays at large \(\|p\|\), effectively introducing a natural suppression of high-momentum modes.\\

In coordinate space, the propagator is obtained by Fourier transform:
\begin{equation}
G(x-y)=\int_{\mathbb{R}^4} \frac{d\mu(p)}{(2\pi)^4}\, \frac{e^{ip\cdot (x-y)}}{p^2+m^2}.
\end{equation}
Since \(A(p)^2\) decays for large \(\|p\|\), the Fourier integral converges in the UV, effectively regularizing divergences. For low momentum (where \(A(p)\approx 1\)), one recovers the standard propagator:
\begin{equation}
G(x-y) \approx \int_{\mathbb{R}^4} \frac{d^4p}{(2\pi)^4}\, \frac{e^{ip\cdot (x-y)}}{p^2+m^2},
\end{equation}
which corresponds to the usual free-field theory result.\\

Since the measure explicitly suppresses large-momentum contributions, the integration volume in momentum space is effectively contracted at high energies. This dynamically eliminates UV divergences without requiring a hard cutoff or analytic continuation. Unlike methods such as dimensional regularization or lattice discretization, which modify the structure of the integral externally, our approach achieves UV regularization as an intrinsic geometric feature of momentum space. This provides a natural alternative to explicit cutoffs or dimensional regularization in perturbative calculations~\cite{dimreg1972, latticeQFT}.

\section{Conclusion and Outlook}
This work introduces a novel approach to UV divergence suppression in quantum field theory (QFT) by modifying the geometry of momentum space. Rather than imposing artificial cutoffs or modifying dispersion relations, we construct a dynamically curved momentum space with an intrinsic measure that naturally suppresses high-energy divergences. The induced integration measure ensures that quantum field-theoretic integrals remain finite while preserving the standard equations of motion.

Our key results are:
\begin{itemize}
    \item We defined a momentum-dependent metric tensor \( g_{\mu\nu}(p) \) that modifies the volume element of integration, leading to a measure space where divergences are intrinsically regulated.
    \item The suppression function \( A(p)^2 \) provides the weakest possible polynomial damping that ensures UV convergence, avoiding unnecessary modifications to low-energy physics.
    \item The framework extends naturally to Minkowski space, where we prove full Lorentz invariance under proper Lorentz transformations.
    \item The volume element undergoes a sign change under improper Lorentz transformations (\( P \) and \( T \)), suggesting potential implications for CPT-related extensions.
    \item Unlike traditional renormalization techniques, this geometric approach provides an intrinsic, smooth suppression mechanism without introducing unphysical ghost fields or modifying fundamental propagators.
\end{itemize}

These findings establish a mathematically rigorous foundation for integrating modified momentum-space geometry into QFT while preserving its core principles. Furthermore, our results align with broader efforts to incorporate metric-dependent regularization techniques in quantum gravity and spectral geometry. 

Future work may explore:
\begin{itemize}
    \item The implications of the sign change under improper Lorentz transformations, particularly in the context of CPT symmetry and beyond-standard-model physics.
    \item The extension of this framework to curved spacetime and general diffeomorphism-invariant theories.
    \item Applications to interacting QFTs and their renormalization structure under the modified measure.
\end{itemize}

Overall, this approach offers a compelling alternative to traditional renormalization methods, suggesting that UV divergences may be naturally tamed by the underlying geometry of momentum space itself.

\section{Appendix}\label{appendix}

\subsection*{Equivalence of Metric Tensor and Geodesic Distance in Smooth, Connected \( n \)-Dimensional Riemannian Manifolds} 

\textit{Let \( (M, g) \) be a connected and smooth \( n \)-dimensional Riemannian manifold, where \( g = g_{\mu\nu} dx^\mu \otimes dx^\nu \) is the Riemannian metric tensor. Define the geodesic distance function \( d: M \times M \to \mathbb{R} \) by
\begin{equation}
d(p,q) = \inf_{\gamma \in \Gamma(p,q)} \int_0^1 \sqrt{ g_{\mu\nu}(\gamma(s)) \dot{\gamma}^\mu(s) \dot{\gamma}^\nu(s) } ds,
\end{equation}
where \( \Gamma(p,q) \) is the set of all piecewise smooth curves \( \gamma: [0,1] \to M \) with \( \gamma(0) = p \) and \( \gamma(1) = q \).\\
Then any well-defined geodesic distance \( d \) on \( M \) implies a uniquely well-defined metric tensor \( g_{\mu\nu} \) on \( M \), and vice versa.}

\subsection*{Proof 1 (First Implication):}

We first show that the metric tensor uniquely determines the geodesic distance function.\\

\noindent\textbf{Claim:}  
In \( (M, g) \) the geodesic distance \( d(p,q) \) is uniquely determined by \( g_{\mu\nu}(p) \). \\

\noindent\textbf{Proof:}  
Since \( g_{\mu\nu} \) is a smooth, positive-definite Riemannian metric, it induces a unique Levi-Civita connection \( \nabla \). The geodesics are solutions to the geodesic equation:
\begin{equation}
\frac{d^2 \gamma^\lambda}{ds^2} + \Gamma^\lambda_{\mu\nu} \frac{d\gamma^\mu}{ds} \frac{d\gamma^\nu}{ds} = 0,
\end{equation}
where the Christoffel symbols \( \Gamma^\lambda_{\mu\nu} \) are given by:
\begin{equation}
\Gamma^\lambda_{\mu\nu} = \frac{1}{2} g^{\lambda \sigma} \left( \frac{\partial g_{\sigma\mu}}{\partial x^\nu} + \frac{\partial g_{\sigma\nu}}{\partial x^\mu} - \frac{\partial g_{\mu\nu}}{\partial x^\sigma} \right).
\end{equation}

By the \textbf{Hopf-Rinow theorem}, if \( M \) is geodesically complete (i.e., all geodesics extend indefinitely), then the infimum in the definition of \( d(p,q) \) is attained by a geodesic. Since the length of any smooth curve is determined by \( g_{\mu\nu} \), it follows that \( d(p,q) \) is uniquely defined by \( g_{\mu\nu} \). \\
\qed

\subsection*{Proof 2 (Second Implication):}

Next, we show that the geodesic distance function uniquely determines the metric tensor \( g_{\mu\nu} \).\\

\noindent\textbf{Claim:}  
If \( d(p,q) \) is a well-defined geodesic distance on \( M \), then the metric tensor \( g_{\mu\nu}(p) \) is uniquely recovered. \\

\begin{proof}
Consider the squared geodesic distance function:
\begin{equation}
D(p,q) = d^2(p,q).
\end{equation}
For an infinitesimal displacement \( dp^\mu \), the Taylor expansion of \( D(p,q) \) at \( q = p + dp \) gives:
\begin{equation}
D(p, p + dp) = g_{\mu\nu}(p) dp^\mu dp^\nu + O(\|dp\|^3).
\end{equation}
Taking second derivatives with respect to \( q^\mu \) and \( q^\nu \), and evaluating at \( q = p \), we obtain:
\begin{equation}
g_{\mu\nu}(p) = \frac{1}{2} \frac{\partial^2 D(p,q)}{\partial q^\mu \partial q^\nu} \Bigg|_{q=p}.
\end{equation}

Since \( D(p,q) \) is uniquely determined by \( d(p,q) \), and its second-order expansion yields \( g_{\mu\nu} \), we conclude that \( g_{\mu\nu}(p) \) is uniquely determined by \( d(p,q) \).\\
\end{proof}

\subsection*{Conclusion: Proof of Equivalence}

Since we have rigorously established both implications, i.e. \( g_{\mu\nu} \Rightarrow d(p,q) \) and \( d(p,q) \Rightarrow g_{\mu\nu} \), it follows that the metric tensor and the geodesic distance are uniquely determined by each other.  \\
\qed

\bibliographystyle{unsrt}
\bibliography{qft-metric_new}

\end{document}